\documentclass[twocolumn,noshowpacs,aps,prl,superscriptaddress]{revtex4}

\usepackage{graphicx}
\usepackage{dcolumn}
\usepackage{amsmath}
\usepackage{epsfig}

\RequirePackage{xspace}

\usepackage{relsize}
\def\babar{\mbox{\slshape B\kern-0.1em{\smaller A}\kern-0.1em
    B\kern-0.1em{\smaller A\kern-0.2em R}}}

\def\epem       {\ensuremath{e^+e^-}\xspace}

\def\g     {\ensuremath{\gamma}\xspace}

\def\q     {\ensuremath{q}\xspace}

\def\qqbar {\ensuremath{q\overline q}\xspace}

\def\piz   {\ensuremath{\pi^0}\xspace}

\def\pip   {\ensuremath{\pi^+}\xspace}
\def\pim   {\ensuremath{\pi^-}\xspace}
\def\pipi  {\ensuremath{\pi^+\pi^-}\xspace}

\def\Kbar  {\kern 0.2em\overline{\kern -0.2em K}{}\xspace}

\def\Kz    {\ensuremath{K^0}\xspace}
\def\Kzb   {\ensuremath{\Kbar^0}\xspace}
\def\KzKzb {\ensuremath{\Kz \kern -0.16em \Kzb}\xspace}
\def\Kp    {\ensuremath{K^+}\xspace}
\def\Km    {\ensuremath{K^-}\xspace}

\def\KpKm  {\ensuremath{\Kp \kern -0.16em \Km}\xspace}
\def\KS    {\ensuremath{K^0_{\scriptscriptstyle S}}\xspace} 
 
\def\Kstarz  {\ensuremath{K^{*0}}\xspace}

\def\Kstar   {\ensuremath{K^*}\xspace}

\def\Kstarp  {\ensuremath{K^{*+}}\xspace}

\def\Dbar    {\kern 0.2em\overline{\kern -0.2em D}{}\xspace}

\def\Dz      {\ensuremath{D^0}\xspace}
\def\Dzb     {\ensuremath{\Dbar^0}\xspace}
\def\DzDzb   {\ensuremath{\Dz {\kern -0.16em \Dzb}}\xspace}
\def\Dp      {\ensuremath{D^+}\xspace}
\def\Dm      {\ensuremath{D^-}\xspace}

\def\DpDm    {\ensuremath{\Dp {\kern -0.16em \Dm}}\xspace}

\def\B       {\ensuremath{B}\xspace}
\def\Bbar    {\kern 0.18em\overline{\kern -0.18em B}{}\xspace}

\def\BB      {\ensuremath{B\Bbar}\xspace} 
\def\Bz      {\ensuremath{B^0}\xspace}
\def\Bzb     {\ensuremath{\Bbar^0}\xspace}
\def\BzBzb   {\ensuremath{\Bz {\kern -0.16em \Bzb}}\xspace}
\def\Bu      {\ensuremath{B^+}\xspace}
\def\Bub     {\ensuremath{B^-}\xspace}
\def\Bp      {\ensuremath{\Bu}\xspace}

\def\BpBm    {\ensuremath{\Bu {\kern -0.16em \Bub}}\xspace}

\def\BorBbar    {\kern 0.18em\optbar{\kern -0.18em B}{}\xspace}
\def\DorDbar    {\kern 0.18em\optbar{\kern -0.18em D}{}\xspace}
\def\KorKbar    {\kern 0.18em\optbar{\kern -0.18em K}{}\xspace}

\def\chiczero {\ensuremath{\chi_{c0}}\xspace}

\mathchardef\Upsilon="7107
\def\Y#1S{\ensuremath{\Upsilon{(#1S)}}\xspace}

\def\FourS {\Y4S}

\mathchardef\Deltares="7101
\mathchardef\Xi="7104
\mathchardef\Lambda="7103
\mathchardef\Sigma="7106
\mathchardef\Omega="710A

\def\Deltabar{\kern 0.25em\overline{\kern -0.25em \Deltares}{}\xspace}
\def\Lbar{\kern 0.2em\overline{\kern -0.2em\Lambda\kern 0.05em}\kern-0.05em{}\xspace}
\def\Sigbar{\kern 0.2em\overline{\kern -0.2em \Sigma}{}\xspace}
\def\Xibar{\kern 0.2em\overline{\kern -0.2em \Xi}{}\xspace}
\def\Obar{\kern 0.2em\overline{\kern -0.2em \Omega}{}\xspace}
\def\Nbar{\kern 0.2em\overline{\kern -0.2em N}{}\xspace}
\def\Xb{\kern 0.2em\overline{\kern -0.2em X}{}\xspace}

\def\mes        {\mbox{$m_{\rm ES}$}\xspace}

\def\DeltaE     {\mbox{$\Delta E$}\xspace}

\newcommand{\tev}{\ensuremath{\mathrm{\,Te\kern -0.1em V}}\xspace}
\newcommand{\gev}{\ensuremath{\mathrm{\,Ge\kern -0.1em V}}\xspace}
\newcommand{\mev}{\ensuremath{\mathrm{\,Me\kern -0.1em V}}\xspace}
\newcommand{\kev}{\ensuremath{\mathrm{\,ke\kern -0.1em V}}\xspace}
\newcommand{\ev}{\ensuremath{\mathrm{\,e\kern -0.1em V}}\xspace}
\newcommand{\gevc}{\ensuremath{{\mathrm{\,Ge\kern -0.1em V\!/}c}}\xspace}
\newcommand{\mevc}{\ensuremath{{\mathrm{\,Me\kern -0.1em V\!/}c}}\xspace}
\newcommand{\gevcc}{\ensuremath{{\mathrm{\,Ge\kern -0.1em V\!/}c^2}}\xspace}
\newcommand{\mevcc}{\ensuremath{{\mathrm{\,Me\kern -0.1em V\!/}c^2}}\xspace}

\def\invfb   {\ensuremath{\mbox{\,fb}^{-1}}\xspace}

\def\mus  {\ensuremath{\rm \,\mus}\xspace}

\def\mus        {\ensuremath{\,\mu{\rm s}}\xspace}    

\def\to                 {\ensuremath{\rightarrow}\xspace}

\def\pep2{PEP-II}

\newcommand{\chisq}{\ensuremath{\chi^2}\xspace}

\def\gsim{{~\raise.15em\hbox{$>$}\kern-.85em
          \lower.35em\hbox{$\sim$}~}\xspace}
\def\lsim{{~\raise.15em\hbox{$<$}\kern-.85em
          \lower.35em\hbox{$\sim$}~}\xspace}

\xspace

\newcommand{\tabref}[1]{Table~\ref{tab:#1}}

\def\jetset74   {\mbox{\tt Jetset \hspace{-0.5em}7.\hspace{-0.2em}4}\xspace}

\newcommand{\rhop}               {\mbox{$\rho^+$}}

\newcommand{\fz}                 {\mbox{$f_0$}}
\newcommand{\fI}                 {\mbox{$\fz(980)$}}
\newcommand{\fII}                {\mbox{$f_2(1270)$}}

\def\fscf	{\mbox{$f_{\rm SCF}$}\xspace}

\def\ncand    {\ensuremath{31\,673}}
\def\nsig     {\ensuremath{1220 \pm 85}}
\def\nsigEffCor {\ensuremath{7427 \pm 518}}

\def\nsigma   {\ensuremath{15.6}}
\def\nsigSys  {\ensuremath{10\,\sigma}}
\def\kpipiBFal{\ensuremath{\left(15.5 \pm 1.1 \pm 1.6 \right)\times 10^{-6}}}

\newcommand{\onreslumi}  {\mbox{429\invfb}}
\newcommand{\offreslumi} {\mbox{45\invfb}}
\newcommand{\bbpairs}    {\mbox{$470.9\pm2.8$~million}}

\newcommand{\nbb}        {\mbox{$N_{\BB}$}}
\newcommand{\NN}         {\mbox{${\rm NN}_{\rm out}$}}

\newcommand{\splot}    {\mbox{$_s{\cal P}lot$}\xspace}
\newcommand{\sweights} {\mbox{$_s{\cal W}eights$}\xspace}

\newcommand{\BABARPubYear}    {10}
\newcommand{\BABARPubNumber}  {001}
\newcommand{\SLACPubNumber} {14108}

\begin{document}

\preprint{\babar-PUB-\BABARPubYear/\BABARPubNumber} 
\preprint{SLAC-PUB-\SLACPubNumber} 

\begin{flushleft}
\babar-CONF-\BABARPubYear/\BABARPubNumber\\
SLAC-PUB-\SLACPubNumber\\
\end{flushleft}

\title{
  {\large 
    \bf Observation of the Rare Decay {\boldmath $\Bp\to\Kp\piz\piz$}
  }
}

%
\author{P.~del~Amo~Sanchez}
\author{J.~P.~Lees}
\author{V.~Poireau}
\author{E.~Prencipe}
\author{V.~Tisserand}
\affiliation{Laboratoire d'Annecy-le-Vieux de Physique des Particules (LAPP), Universit\'e de Savoie, CNRS/IN2P3,  F-74941 Annecy-Le-Vieux, France}
\author{J.~Garra~Tico}
\author{E.~Grauges}
\affiliation{Universitat de Barcelona, Facultat de Fisica, Departament ECM, E-08028 Barcelona, Spain }
\author{M.~Martinelli$^{ab}$}
\author{A.~Palano$^{ab}$ }
\author{M.~Pappagallo$^{ab}$ }
\affiliation{INFN Sezione di Bari$^{a}$; Dipartimento di Fisica, Universit\`a di Bari$^{b}$, I-70126 Bari, Italy }
% ...
\collaboration{The \babar\ Collaboration}
\noaffiliation

\date{\today}

\begin{abstract}
We report an analysis of charmless
hadronic decays of charged \B\ mesons to the final state $\Kp\piz\piz$,
using a data sample of \bbpairs\ \BB\ events collected with the
\babar\ detector at the \FourS\ resonance.
We observe an excess of signal events with a significance above
10 standard deviations including systematic uncertainties and measure
the branching fraction to be 
${\cal B}\left(\Bp\to\Kp\piz\piz\right) = \kpipiBFal$, 
where the uncertainties are statistical and systematic, respectively. 
\end{abstract}


\maketitle


Recent measurements of rates and asymmetries in $B\to K\pi$ decays have
have generated considerable interest because of possible hints of new physics
contributions~\cite{Aubert:2007hh,:2008zza}. 
Unfortunately, hadronic uncertainties prevent a clear interpretation of these
results in terms of physics beyond the Standard Model (SM).
A data driven approach, involving measurements of all observables in the $B\to
K\pi$ system can in principle resolve the theoretical situation, but much more
precise measurements ({\it i.e.} much larger data samples) will be
needed~\cite{Fleischer:2008wb,Gronau:2008gu,Ciuchini:2008eh}. 

It is interesting to study the related decays to pseudoscalar-vector
final states $B\to \Kstar\pi$ and $B\to
K\rho$~\cite{Chang:2008tf,Chiang:2009hd,Gronau:2010dd}. 
In \tabref{status} we review the existing experimental measurements of the
channels in the $B\to \Kstar\pi$ system. 
It is evident that improved measurements of the $\Kstarp\piz$~\cite{cc} decay
are needed.

\begin{table}[htb]
\center
\caption{
  Experimental measurements of $B\to \Kstar\pi$ decays.
  Average values come from HFAG~\cite{Barberio:2008fa}.
}
\label{tab:status}
\begin{tabular}{c@{\hspace{5mm}}c@{\hspace{5mm}}c@{\hspace{5mm}}c}
\hline
Mode & ${\cal B}\times10^{6}$ & $A_{CP}$ & References \\
\hline
$\Kstarp\pim$ & $10.3 \pm 1.1$ & $-0.23 \pm 0.08$ & \cite{Aubert:2007bs,Aubert:2008zu,Garmash:2006fh,:2008wwa} \\
$\Kstarp\piz$ & $6.9 \pm 2.3$ & $0.04 \pm 0.29 \pm 0.05$ & \cite{Aubert:2005cp} \\
$\Kstarz\pip$ & $9.9\,^{+0.8}_{-0.9}$ & $-0.020\,^{+0.067}_{-0.061}$ &
\cite{Aubert:2008bj,Garmash:2005rv,Belle:Kpipi} \\
$\Kstarz\piz$ & $2.4 \pm 0.7$ & $-0.15 \pm 0.12 \pm 0.02$ & \cite{Chang:2004um,Aubert:2008zu} \\
\hline
\end{tabular}
\end{table}

Due to the non-negligible width of the \Kstar\ resonances, the quasi-two-body
modes are best studied in the analysis of the three-body Dalitz plot.
The four $\Kstar\pi$ decays populate six $K\pi\pi$ Dalitz plots (the four
$K\rho$ decays also produce four of the same six final states).
Of these, Dalitz plot analyses of
$\Kp\pipi$~\cite{Aubert:2008bj,Garmash:2005rv},
$\KS\pipi$~\cite{Aubert:2009me,:2008wwa} and
$\Kp\pim\piz$~\cite{Aubert:2008zu} have been performed to date. 
The first two of these have shown the presence of a poorly-understood
structure, dubbed the $f_{\rm X}(1300)$, in the $\pipi$ invariant mass
distribution. 
A study of the $\Kp\piz\piz$ Dalitz plot would help to elucidate the nature of
this peak, since even-spin states will populate both $K\pipi$ and $K\piz\piz$
(assuming isospin symmetry), while odd-spin states cannot decay to $\piz\piz$.

Knowledge of the $\Kp\piz\piz$ Dalitz plot may also help to clarify the interpretation of
the inclusive time-dependent analyses~\cite{Gershon:2004tk} of
$\Bz\to\KS\piz\piz$~\cite{Aubert:2007ub,:2007xd}.  
Currently, these results show the largest deviation, albeit with a large
uncertainty, among hadronic $b \to s$ penguin-dominated
decays~\cite{Barberio:2008fa} 
from the na\"ive Standard Model expectation that the time-dependent $CP$
violation parameter should be given by $S_{CP} \approx -\eta_{CP}
\sin(2\beta)$, where $\eta_{CP}$ is the $CP$ eigenvalue of the final state
($+1$ for $\KS\piz\piz$) and $\beta$ is an angle of the
Cabibbo-Kobayashi-Maskawa~\cite{Cabibbo:1963yz,Kobayashi:1973fv} 
unitarity triangle.
Such deviations could be caused by new physics, but in order to rule out the
possibility of sizeable corrections to the Standard Model prediction,
better understanding of the population of the $K\piz\piz$ Dalitz plots will be
necessary.

In this article, we present the results of a search for the three-body decay
$\Bp\to\Kp\piz\piz$, including short-lived intermediate two-body modes that
decay to this final state.
This is the first step towards measuring the properties of contributing
resonant modes.
There is no existing previous measurement of the three-body branching
fraction, but several quasi-two-body modes that can decay to this final state
have been seen, with varying significances.
These include $\Bp\to\fI\Kp$, observed in the
$\fI\to\pipi$ channel~\cite{Aubert:2008bj,Garmash:2005rv} and also seen in
$\fI\to\Kp\Km$~\cite{Aubert:2006nu},
$\Bp\to\fII\Kp$, seen in $\fII\to\pipi$~\cite{Aubert:2008bj,Garmash:2005rv},
and $\Bp\to\Kstarp(892)\piz$, seen in
$\Kstarp(892)\to\Kp\piz$~\cite{Aubert:2005cp}.
The decay $\Bp\to\chiczero\Kp$ has also been
observed with $\chiczero\to\pipi$~\cite{Aubert:2008bj,Garmash:2005rv}
and $\chiczero\to\Kp\Km$~\cite{Aubert:2006nu,Garmash:2004wa}.

The data used in the analysis, collected with the
\babar\ detector~\cite{Aubert:2001tu} at the \pep2\ asymmetric energy 
\epem\ collider at SLAC, consist of an integrated luminosity of \onreslumi\
recorded at the \FourS\ resonance (``on-peak'') and \offreslumi\ collected
40\,\mev\ below the resonance (``off-peak''). 
The on-peak data sample contains the whole \babar\ dataset of \bbpairs\ \BB\ events.

We reconstruct $\Bp\to\Kp\piz\piz$ decay candidates by 
combining a \Kp\ candidate with two neutral pion candidates.
The \Kp\ candidates are required to have a minimum
transverse momentum of 0.05\,\gevc\ and to be consistent with having originated
from the interaction region.  Separation of charged kaons from charged pions
is accomplished with energy-loss information from the tracking subdetectors,
and the Cherenkov angle and number of photons measured by a
ring-imaging Cherenkov detector.
The efficiency for kaon selection is approximately 80\,\% including
geometrical acceptance, while the probability of misidentification of pions
as kaons is below 5\,\% up to a laboratory momentum of 4\,\gevc. 
Neutral pion candidates are formed from pairs of photons with laboratory
energies above $0.05 \gev$ and lateral moments between $0.01$ and $0.6$.  
We require that the mass of the reconstructed \piz\ is
within the range $0.115 \gevcc < m_{\g\g} <0.150 \gevcc$ and that the
absolute value of the cosine of the decay angle in the \piz\ rest frame is
less than $0.9$.  
We exclude candidates that are consistent with the $\Bp\to\KS\Kp$,
$\KS\to\piz\piz$ decay chain by rejecting events that contain a candidate that
satisfies $0.40 \gevcc < m_{\piz\piz} < 0.55 \gevcc$.
This veto has a signal efficiency of at least $96\,\%$ for any charmless
resonant decay and is almost $100\,\%$ efficient for nonresonant
$\Bp\to\Kp\piz\piz$ and $\Bp\to\chiczero\Kp$ decays.

Due to the presence of two neutral pions in the final state, there is a
significant probability for signal events to be misreconstructed, 
due to low momentum particles being exchanged with particles from the decay of
the other \B\ meson in the event.  
We refer to these as ``self-cross-feed'' (SCF) events, as opposed to correctly
reconstructed (CR) events.
Using a classification based on Monte Carlo information, we find that in
simulated events the SCF fraction depends strongly on the Dalitz plot
distribution of the signal, and ranges from $2\,\%$ for $\Bp\to\chiczero\Kp$
decays to $30\,\%$ for $\Bp\to\fII\Kp$ decays.

To suppress the dominant background contribution, which arises from continuum
$\epem\to\qqbar\ (\q=u,d,s,c)$ events, we employ a neural network that
combines four variables commonly used to
discriminate jet-like \qqbar\ events from the more spherical \BB\ events. 
These are the ratio of the second to the zeroth order momentum-weighted
angular moment, the absolute value of the cosine of the angle between the 
\B\ direction and the beam axis, the absolute value of the cosine 
of the angle between the \B\ thrust axis and the beam axis, and the absolute
value of the output of a neural network used for ``flavour tagging'' 
({\it i.e.} for distinguishing \B\ from $\Bbar$ decays using inclusive
properties of the decay of the other \B\ meson in the $\FourS\to\B\Bbar$
decay~\cite{Aubert:2009yr}). 
The first three quantities are calculated in the center-of-mass (CM) frame.
We apply a loose criterion on the neural network output (\NN) 
which retains approximately $90\,\%$ of the signal while rejecting
approximately $82\,\%$ of the $\qqbar$ background.

In addition to \NN, we distinguish signal from background events using two
kinematic variables: the difference \DeltaE\ between the CM energy of the
\B\ candidate and $\sqrt{s}/2$, and the beam-energy-substituted mass
$\mes=\sqrt{s/4-{\bf p}^2_\B}$, where $\sqrt{s}$ is the total CM energy and
${\bf p}_\B$ is the momentum of the candidate \B\ meson in the CM frame.
The signal \mes\ distribution for CR events is approximately independent of
the $\Bp\to\Kp\piz\piz$ Dalitz plot distribution and peaks near the \B\ mass
with a resolution of about $3\mevcc$.
We select signal candidates that satisfy $5.260 \gevcc < \mes < 5.286 \gevcc$.
The CR signal \DeltaE\ distribution peaks near zero, but has a resolution that
depends on the signal Dalitz plot distribution, which is {\it a priori} unknown.
To avoid possible biases~\cite{Punzi:2004wh} we apply tighter selection
criteria, $-0.15 \gev < \DeltaE < 0.05 \gev$, which have an efficiency of
about $80\,\%$ for signal while retaining only about $30\,\%$ of the
background (both compared to the looser requirement $|\DeltaE|<0.30\gev$), and
do not use \DeltaE\ in the fit described below.

The efficiency for signal events to pass all the selection criteria is
determined as a function of position in the Dalitz plot.
Using a Monte Carlo (MC) simulation in which events uniformly populate the
phase-space, we obtain an average efficiency of approximately $16\,\%$, though
values as low as $8\,\%$ are found near the corners of the Dalitz plot.

An average of $1.3$ \B\ candidates is found per selected event.
In events with multiple candidates we choose the one with the smallest value
of a \chisq\ variable formed from the sum of the \chisq\ values of the two
\piz\ candidate masses.
This procedure has been found to select the best reconstructed candidate more
than $90\,\%$ of the time, and does not bias our fit variables.

We study residual background contributions from \BB\ events using MC
simulations. 
It is found that these events can be combined
into four categories based on their shapes in \mes\ and \DeltaE.
The first category comprises two-body modes (mainly $\Bp\to\Kp\piz$);
the second contains three-body modes 
(mainly $\Bp\to\Kstarp\gamma$ and $\Bp\to\pip\piz\piz$);
the third and fourth are composed of higher multiplicity decays (many possible
sources, with or without intermediate charmed states) with missing particles,
and are distinguished by the absence or presence of a peak in the 
\mes\ distribution respectively. 
Based on the MC-derived efficiencies, total number of \BB\ events, and known
branching fractions~\cite{Barberio:2008fa,Amsler:2008zzb}, 
we expect $70 \pm 9$, $39 \pm 18$, $1090 \pm 40$ and $170 \pm 30$ events in
the four respective categories. 

To obtain the $\Bp\to\Kp\piz\piz$ signal yield, we perform an
unbinned extended maximum likelihood fit to the candidate events using
two input variables: \mes and \NN.
For each component $j$ (signal, \qqbar\ background, and
the four \BB\ background categories), we define a probability density function
(PDF) 
\begin{equation}
  \label{eq:PDF-exp}
  {\cal P}^i_j \equiv
  {\cal P}_j(\mes^i){\cal P}_j({\NN}^i),
\end{equation}
where $i$ denotes the event index. 
The signal component is further separated into CR and SCF parts
\begin{equation}
  \label{eq:PDF-sig}
  \begin{array}{rcl}
    {\cal P}^i_{\rm sig} & \equiv &  
    (1 - \fscf) {\cal P}_{\rm CR}(\mes^i){\cal
      P}_{\rm CR}({\cal \NN}^i) + \\
    & \multicolumn{2}{r}{\fscf {\cal P}_{\rm SCF}(\mes^i) 
       {\cal P}_{\rm SCF}({\cal \NN}^i)\, ,}
  \end{array}
\end{equation}
where \fscf\ is the SCF fraction.
The extended likelihood function is
\begin{equation}
  \label{eq:extML-Eq}
  {\cal L} =
  \prod_{k} e^{-n_k}
  \prod_{i}\left[ \sum_{j}n_j{\cal P}^i_j \right],
\end{equation}
where $n_{j(k)}$ is the yield of the event category $j(k)$.

For the signal, the \mes\ PDFs for CR and SCF are described by an
asymmetric Gaussian with power-law tails and a third order Chebychev
polynomial, respectively.
Both CR and SCF \NN\ PDFs are described by one-dimensional histograms.
We fix the shape parameters to the
values obtained from the $\Bp\to\Kp\piz\piz$ phase-space MC sample,
after adjusting them to account for possible differences between data
and MC simulations determined with a high statistics control sample of
$\Bp\to\Dzb\rhop\to\left(\Kp\pim\piz\right)\left(\pip\piz\right)$ decays. 
For the continuum background, we use an ARGUS
function~\cite{Albrecht:1990am} to parameterize the \mes\ shape.
The continuum \NN\ shape is modelled with a parametric step function function
with 20 bins.
One-dimensional histograms are used as nonparametric PDFs to represent all fit
variables for the four \BB\ background components.
The free parameters of our fit are the yields of signal and continuum
background together with the parameters of the continuum \mes\ and \NN\ PDFs. 
All the yields and PDF shapes of the four $\BB$ background categories are
fixed based on MC simulations. 

The results of the fit are highly sensitive to the value of \fscf, which
depends strongly on the Dalitz plot distribution of signal events and cannot
be determined directly from the fit.
To circumvent this problem, we adopt an iterative procedure.
We perform a fit with \fscf\ fixed to an initial value.
We then construct the signal Dalitz plot from the signal probabilities for
each candidate event (\sweights) calculated with the
\splot\ technique~\cite{Pivk:2004ty}, and determine the corresponding
average value of \fscf.
We then fit again with \fscf\ fixed to the new value, and repeat until
the obtained values of the total signal yield (CR + SCF) and \fscf\ are
unchanged between iterations.
This method was validated using MC and was found to return values of 
\fscf\ that are accurate to within $3\,\%$ of the nominal SCF fraction.
Convergence is typically obtained within 3 iterations.

We cross-check our analysis procedure using the high statistics control
sample described above.
We impose selection requirements on the $D$ and $\rho$ candidates' invariant
masses: $1.84 \gevcc <m_{\Kp\pim\piz}< 1.88 \gevcc$ and 
$0.65 \gevcc <m_{\pip\piz}< 0.85 \gevcc$.
We fit the data with a likelihood function that includes components for the
control channel, and for backgrounds from \BB\ and \qqbar\ backgrounds.
We find a yield consistent  within statistical uncertainties with the
expectation based on world-average product branching
fractions~\cite{Amsler:2008zzb}. 

\begin{figure*}[!htb]
 \includegraphics[width=.4942\textwidth]{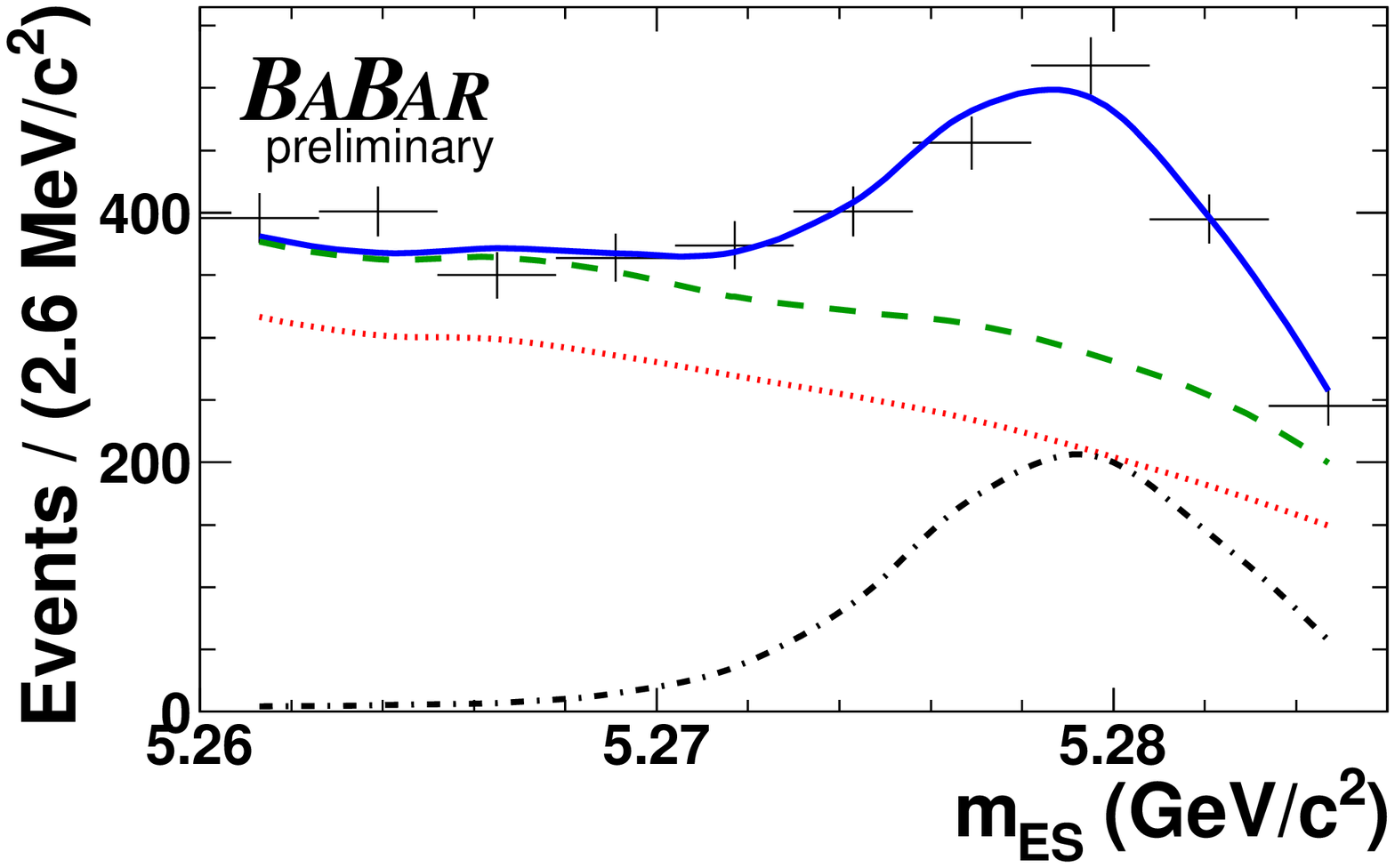} 
 \includegraphics[width=.4942\textwidth]{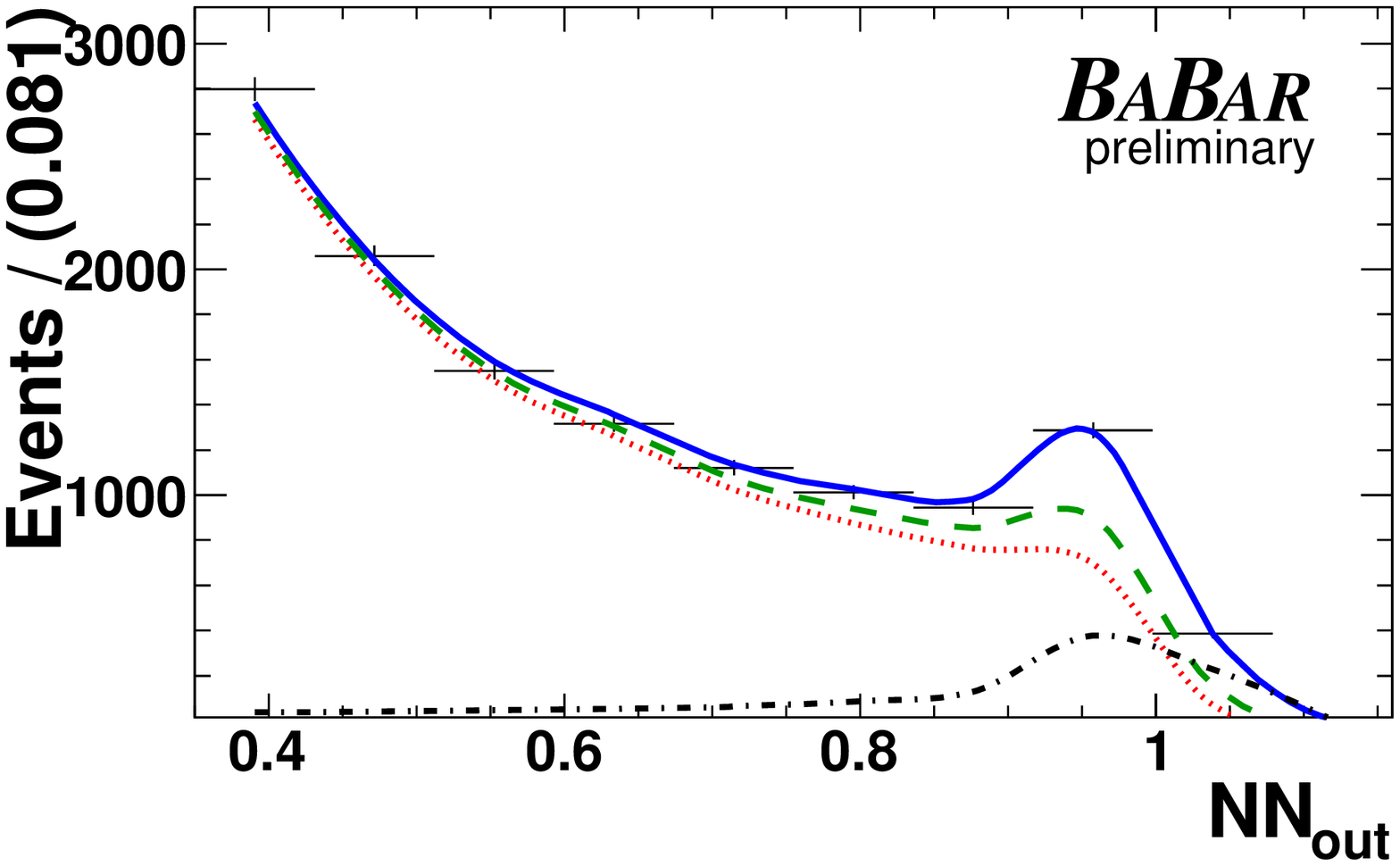}
 \caption{
   Projections of candidate events onto \mes\ (left) and \NN (right),
   following requirements on the other fit variable in order to enhance signal
   visibility. 
   Points with error bars show the data, the solid (blue) curves represent the
   total fit result, the dashed (green) curves show the total background  
   contribution, and the dotted (red) curves are the \qqbar\ component. The
   dash-dotted curves represent the signal contribution.
 }
 \label{fig:signal-project}
 \end{figure*}

We apply the fit method described above to the \ncand\ selected candidate
$\Bp\to\Kp\piz\piz$ events.  Convergence is obtained after four iterations
with a yield of \nsig\ signal events and a SCF fraction of $9.7\,\%$.
The results of the fit are shown in Fig.~\ref{fig:signal-project}.
The statistical significance of the signal yield, 
given by $\sqrt{2\Delta\ln{\cal L}}$ where $\Delta\ln{\cal L}$ is the
difference between the
negative log likelihood obtained assuming zero signal events and that at its
minimum, is \nsigma\ standard deviations ($\sigma$).
Including systematic uncertainties (discussed below), the significance is
above \nsigSys.

The $\Bp\to\Kp\piz\piz$ branching fraction is determined from the result of
the fit, dividing the signal \sweights\ by event-by-event efficiencies that
take the Dalitz-plot position dependence into account, and summing them to
obtain an efficiency-corrected signal yield of \nsigEffCor\ events.
We further correct for the effect of the \KS\ veto and a bias in the fitted 
signal yield, as determined from
Monte Carlo pseudoexperiments generated with a signal component with the same
values of the yield and SCF fraction as found in the fit to data.
Finally, we divide by the total number of \BB\ events in the data sample
to obtain our measurement of the branching fraction
${\cal B}\left(\Bp\to\Kp\piz\piz\right) = \kpipiBFal$,
where the first uncertainty is statistical and the second is systematic.

We assign systematic uncertainties due to 
(i)
uncertainties in CR signal \mes\ PDF shapes (0.8\,\%) evaluated using the
$\Bp\to\Dzb\rhop\to\left(\Kp\pim\piz\right)\left(\pip\piz\right)$ control
sample;
(ii)
uncertainties in CR signal and \BB\ background \NN\ PDF shapes (4.9\,\%)
evaluated using uncertainties in the data/MC ratio determined from the
control sample and applying them to all PDFs in a correlated manner;
(iii)
uncertainties in the SCF signal \mes\ and \NN\ PDF shapes (1.7\,\% and
0.7\,\%, respectively) evaluated considering a range of SCF shapes
corresponding to different signal Dalitz plot distributions;
(iv)
uncertainty in the SCF fraction (2.5\,\%) from varying the value used in
the fit within a range of uncertainty determined from Monte Carlo
pseudoexperiment tests of our iterative fitting procedure;
(v)
uncertainties in the \BB\ background PDFs due to finite MC statistics
(0.8\,\%), determined by varying the contents of the bins of the histograms
used to describe the PDFs within their errors;
(vi)
uncertainties in the \BB\ background \mes\ PDF shapes due to data/MC
differences (1.6\,\%), evaluated by smearing the PDFs with a Gaussian with
parameters determined from the control sample; 
(vii)
uncertainties in the fixed \BB\ background yields (1.4\,\%), evaluated by
varying these within their uncertainties;
(viii)
uncertainty in the correction due to fit bias (1.8\,\%), which corresponds
to half the correction combined in quadrature with its error;
(ix)
uncertainties in the efficiency, with contributions from tracking
(0.4\,\%), kaon identification (1.0\,\%), neutral pion reconstruction
(3.0\,\% per neutral pion, so 6.0\,\% in total), \DeltaE\ (4.0\,\%) and
\NN\ (3.0\,\%) selection requirements, the \KS\ veto correction (2.0\,\%);
(x)
uncertainty in the number of \BB\ pairs in the data sample (0.6\,\%).
The total systematic uncertainty on the branching fraction is 10.4\,\%.
Including only systematic uncertainties that affect the yield, the total is
6.5\,\%. \tabref{systematics} summarizes the systematic contributions.

\begin{table}
\caption{
  Summary of systematic uncertainties.
}
\label{tab:systematics}
\begin{tabular}{l|c}
\hline
Source & Uncertainty \\
\hline
CR signal \mes\ PDF & 0.8\,\% \\
CR signal and \BB\ background \NN\ PDFs & 4.9\,\% \\
SCF signal \mes\ PDF & 1.7\,\% \\
SCF signal \NN\ PDF & 0.7\,\%\\
SCF fraction & 2.5\,\% \\
\BB\ background PDFs (MC statistics) & 0.8\,\% \\
\BB\ background \mes\ PDFs & 1.6\,\% \\
\BB\ background yields & 1.4\,\% \\
Fit bias & 1.8\,\% \\
\hline
Subtotal & 6.5\,\% \\
\hline
Tracking efficiency & 0.4\,\% \\
Particle identification & 1.0\,\% \\
Neutral pion efficiency & 6.0\,\% \\
\DeltaE\ cut efficiency & 4.0\,\% \\
\NN\ cut efficiency     & 3.0\,\% \\
\KS\ veto & 2.0\,\% \\
\nbb & 0.6\,\% \\
\hline
Total & 10.4\,\% \\
\hline
\end{tabular}
\end{table}

In summary, using the full \babar\ data sample of \onreslumi\ collected at the
$\FourS$ resonance, we observe charmless hadronic decays of charged
$\B$ mesons to the final state $\Kp\piz\piz$.
The signal has a significance above \nsigSys, after taking systematic effects
into account.  
We measure the branching fraction to be 
${\cal B}\left(\Bp\to\Kp\piz\piz\right) = \kpipiBFal$.
This is the first step towards understanding the composition of the Dalitz
plot of this decay and measuring the properties of contributing
quasi-two-body modes.

We are grateful for the 
extraordinary contributions of our \pep2\ colleagues in
achieving the excellent luminosity and machine conditions
that have made this work possible.
The success of this project also relies critically on the 
expertise and dedication of the computing organizations that 
support \babar.
The collaborating institutions wish to thank 
SLAC for its support and the kind hospitality extended to them. 
This work is supported by the
US Department of Energy
and National Science Foundation, the
Natural Sciences and Engineering Research Council (Canada),
the Commissariat \`a l'Energie Atomique and
Institut National de Physique Nucl\'eaire et de Physique des Particules
(France), the
Bundesministerium f\"ur Bildung und Forschung and
Deutsche Forschungsgemeinschaft
(Germany), the
Istituto Nazionale di Fisica Nucleare (Italy),
the Foundation for Fundamental Research on Matter (The Netherlands),
the Research Council of Norway, the
Ministry of Education and Science of the Russian Federation, 
Ministerio de Ciencia e Innovaci\'on (Spain), and the
Science and Technology Facilities Council (United Kingdom).
Individuals have received support from 
the Marie-Curie IEF program (European Union), the A. P. Sloan Foundation (USA) 
and the Binational Science Foundation (USA-Israel).

\bibliography{references}

\begin{thebibliography}{33}
\bibitem[{cc()}]{cc}
\bibinfo{note}{{The inclusion of charge conjugate modes is implied throughout
  this paper.}}
\end{thebibliography}
\bibliographystyle{apsrev}

\end{document}